\documentclass[aps,prl,showpacs,twocolumn,superscriptaddress, floatfix]{revtex4-1}%

\usepackage[utf8]{inputenc}
\usepackage{epsfig}
\usepackage{url}
\usepackage{multirow}
\usepackage{capt-of} 

\usepackage{graphicx}

\usepackage{amsfonts}
\usepackage{amssymb}
\usepackage{amsmath}
\usepackage{verbatim}
\usepackage{float} 
\usepackage{bm} 
\usepackage{qcircuit}

\usepackage{mathtools}

\usepackage{color}

\usepackage{hyperref}
\hypersetup{  colorlinks=true, linkcolor=blue, citecolor=red, urlcolor=blue  }

\usepackage{physics}

\renewcommand{\bf}[1]{\mathbf{#1}}

\renewcommand{\bold}[1]{\boldsymbol{#1}}
\newcommand{\ba}{\begin{eqnarray}}
\newcommand{\ea}{\end{eqnarray}}

\newcommand{\beginsupplement}{%
        \setcounter{table}{0}
        \renewcommand{\thetable}{S\arabic{table}}%
        \setcounter{figure}{0}
        \renewcommand{\thefigure}{S\arabic{figure}}%
        \setcounter{equation}{0}
        \renewcommand{\theequation}{S\arabic{equation}}%
     }   

\usepackage{amsthm}

\theoremstyle{definition}
\newtheorem{definition}{Definition}

\usepackage{txfonts}

\renewcommand{\H}{\mathsf{H}} 
\DeclareMathOperator{\poly}{poly}
\newcommand{\others}{\operatorname{others}}
\newcommand{\fracp}[2]{\left( \frac{#1}{#2} \right)} 
\def\sharpsat{\textsf{\#2-SAT}}
\def\sharpP{\mathsf{\# P}}

\begin{document}

\title{Ubiquitous Complexity of Entanglement Spectra}

\author{Bin Cheng}
\affiliation{Institute for Quantum Science and Engineering, and Department of Physics, Southern University of Science and Technology, Shenzhen 518055, China}


\author{Man-Hong Yung}
\email{yung@sustech.edu.cn}
\affiliation{Institute for Quantum Science and Engineering, and Department of Physics, Southern University of Science and Technology, Shenzhen 518055, China}
\affiliation{Shenzhen Key Laboratory of Quantum Science and Engineering, Southern University of Science and Technology, Shenzhen 518055, China}







\begin{abstract}
In recent years, the entanglement spectra of quantum states have been identified to be highly valuable for improving our understanding on many problems in quantum physics, such as classification of topological phases, symmetry-breaking phases, and eigenstate thermalization, etc. However, it remains a major challenge to fully characterize the entanglement spectrum of a given quantum state. An outstanding problem is whether the difficulty is intrinsically technical or fundamental? Here using the tools in computational complexity, we perform a rigorous analysis to pin down the counting complexity of entanglement spectra of (i) states generated by polynomial-time quantum circuits, (ii) ground states of gapped 5-local Hamiltonians, and (iii) projected entangled-pair states (PEPS). We prove that despite the state complexity, the problems of counting the number of sizable elements in the entanglement spectra all belong to the class $\sharpP$-complete, which is as hard as calculating the partition functions of Ising models. Our result suggests that the absence of an efficient method for solving the problem is fundamental in nature, from the point of view of computational complexity theory. 

\end{abstract}

\maketitle



\textbf{Introduction.---} Quantum entanglement is a unique feature of the quantum information science, leading to many novel non-classical applications such as quantum teleportation~\cite{bennett_teleportation_1993}, quantum computation~\cite{kitaev2002book}, quantum simulation~\cite{Feynman1982,Lloyd1996-simulation}, etc. Moreover, the notion of quantum entanglement has created a great impact on various branches of physics. Particularly, the application of entanglement entropy to condensed-matter physics leads to a whole new paradigm of understanding many-body systems based on the concept of topological order~\cite{kitaev_topological_2006,levin_detecting_2006}, which goes beyond the traditional symmetry-breaking framework. 

On the other hands, entanglement spectrum was proposed by Li and Haldane~\cite{li_entanglement_2008} as a complementary concept of entanglement entropy. More precisely, for any given bipartite quantum state, $\ket{\xi} = \sum_a \sqrt{\lambda_a} \ket{\eta_a} \ket{\psi_a}$ written in the Schmidt-decomposed form, the reduced state is given by $\rho = \sum_a \lambda_a \dyad{\psi_a}$. The structure of entanglement spectrum of $\ket{\xi}$ is defined as the eigenvalue spectrum of the reduced density matrix $\rho$, i.e. the set of eigenvalues or Schmidt coefficients $\{ \lambda_a \}$. 
Since then, the structure of entanglement spectra has led to many applications in many-body physics, such as classification of topological phases~\cite{pollmann_entanglement_2010,thomale_entanglement_gap_2010,cirac_entanglement_2011,chandran_bulk-edge_2011,qi_general_2012,schuch_topological_2013}, symmetry-breaking phases~\cite{poilblanc_entanglement_2010,cirac_entanglement_2011,alba_boundary_2012,kolley_entanglement_2013,Metlitski_entanglement_2011}, eigenstate thermalization~\cite{Garrison_eigenstate_thermalization_2018}, etc. 

However, the problem of characterizing the entanglement spectrum is notoriously challenging. A naive approach would be quantum state tomography~\cite{tomography2004}, but it requires resources scaling exponentially. Recently, there have been several alternative schemes proposed~\cite{pichler_cold_atom_2016,dalmonte_quantum_2018,johri_entanglement_spectroscopy_2017,beverland_spectrum_2018}. However, all of these approaches become inefficient or ineffective when the system size is scaled up. This leads to the question: \textit{is the challenge of characterizing the entanglement spectrum a purely technological problem or a fundamental one?}





Here we focus on a specific setting in determining the structure of entanglement spectra, where we present rigorous results on the computational complexity in counting the number of Schmidt coefficients that are larger than a given threshold value. We shall prove that for all of the following cases, including {(i)} BQPS: states generated by quantum circuits in polynomial time, {(ii)} G5LS: ground states of gapped 5-local Hamiltonians, and {(iii)} PEPS~\cite{verstraete_2004_peps}, the problems of counting the entanglement spectra (denoted as CES) all belong to the complexity class $\sharpP$-complete~\cite{arora_computational_2009}, i.e., same complexity class as evaluating the partition functions of Ising models~\cite{jaeger_1990_ising_partition}. 

The complexity class of $\sharpP$ contains the set of problems counting the number of solutions of $\mathsf{NP}$ problems. The fact that CES is $\sharpP$-complete implies that all of the $\sharpP$ problems can be recasted as problems in the CES, and that CES itself belongs to the class $\sharpP$. On the other hand, the three classes of quantum states under consideration are ordered in terms of increasing complexity in the following sense: every BQPS can be encoded into some G5LS~\cite{kitaev2002book,aharonov_quantum_np_2002}, and every G5LS can be represented by some PEPS~\cite{schuch_PEPS_complexity_2007}. 
In terms of complexity theory, these three states correspond to the complexity classes $\mathsf{BQP}$, $\mathsf{QMA}$~\cite{kitaev2002book} and $\textsf{post-BQP}$~\cite{aaronson_postselection_2004}, respectively, which have the following relation: $\textsf{BQP} \subseteq \textsf{QMA} \subseteq \textsf{post-BQP}$. 
These results suggest that, to some extent, the complexity of CES problem is independent of the complexity of quantum states. 

Our main techniques can be divided into three parts. In part I, we show that counting the ground-state degeneracy (denoted as CGD) of gapped local Hamiltonians is in $\sharpP$. In part II, by treating reduced density matrix as a Hamiltonian,  we prove that counting entanglement spectrum is also in $\sharpP$. In part III, we prove that both problems are $\sharpP$-hard, and thus $\sharpP$-complete. 

To get started, our problem of counting entanglement spectrum can be formally defined as follows.
\begin{definition}[CES: counting entanglement spectrum]
\label{def:counting_entanglement_spectrum}
For a quantum state of $n$ qubits, given (i) an upper bound $\lambda^* := \mathop {\max }\limits_a \left\{ {{\lambda _a}} \right\}$ for $\lambda_a$ and (ii) a `promise gap' $\Delta_\lambda := \lambda^*/\poly(n)$, output the number of Schmidt coefficients $\lambda_a$ above $\Delta_{\lambda}$.
\end{definition}
Here the promise gap captures the notion of counting entanglement spectrum with an accuracy $\Delta_{\lambda}$. In contrast with the local Hamiltonian problem~\cite{kitaev2002book} where the gap scales as an inverse polynomial, here the gap is dependent on $\lambda^*$. In this way, for some entangled states, $\lambda^*$ can also be  exponentially small and so is $\Delta_{\lambda}$.

Note that, in the work of Li and Haldane~\cite{li_entanglement_2008}, the `entanglement Hamiltonian' $\tilde{H}$ of a density matrix $\rho$ is defined by $\rho = e^{- \tilde{H}}$. In this way, the problem of counting entanglement spectrum can be recasted as counting the eigenstates of $\tilde{H}$ with entanglement energies smaller than a promise gap $-\log{\Delta_{\lambda}}$, i.e. CGD of the entanglement Hamiltonian. 
In Ref.~\cite{browm-density-of-state} and Ref.~\cite{shi_sharp_bqp}, it has already been proved that CGD of gapped local Hamiltonian is $\sharpP$-complete. In this work, we not only provide a novel proof to this result, but also extend it to the case of entanglement Hamiltonian.

\begin{figure}[t]
\centering
\includegraphics[width=0.4\textwidth]{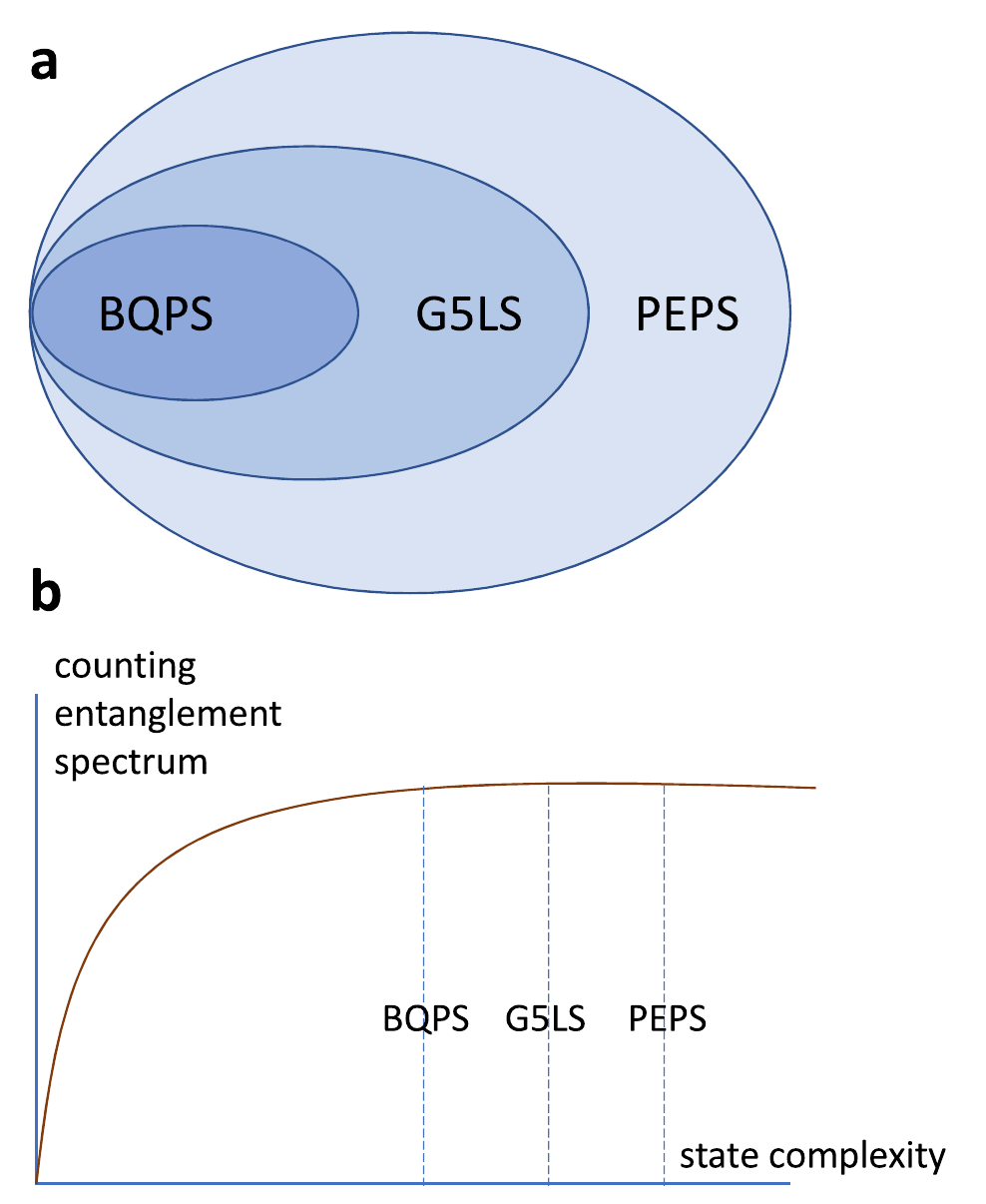}
\caption{\textbf{(a)} The relation between the three kinds of states that we consider in the main text, in terms of complexity theory. \textbf{(b)} Despite the state complexity, the problem of counting entanglement spectra for these three states are equally hard.}
\label{fig:conceptual_relation}
\end{figure}



\textbf{Part I: Counting ground-state degeneracy.---}
Given a local Hamiltonian $H = \sum_i H_i$ with $\| H_i \| \leq 1$ and two real numbers $a$ and $b$, it is promised that $a - b \geq 1/\poly(n)$ and there are no eigenvalues in between. The problem of counting ground-state degeneracy is to count the number of eigenvalues of $H$ below $b$. In this part, we would show that CGD of the Hamiltonian $H$ is in $\sharpP$.

Suppose that $\ket{\psi_i}$ is an eigenstate of $H$ with an eigenvalue $\lambda_i$. To count the ground-state degeneracy, we would need to estimate the value of $\lambda_i$ first, which can be achieved by phase estimation. First, we implement the time evolution $U = e^{-i H t}$ with the truncated-taylor-series method~\cite{berry-truncated-taylor}, and set $t = 2\pi(0, 2^1, \ldots, 2^{d_t})$, where $d_t$ is the significant digit of the binary representation of $\lambda_i$. Then we perform phase estimation:
\ba
\ket{0^{d_t}} \ket{\psi_i} \xrightarrow{\text{phase estimation}} \sum_k q_k^{(i)} \ket{\tilde{\lambda}_k} \ket{\psi_i} \ ,
\ea
where $\abs{q_k^{(i)}}^2$ is the probability of measuring $\ket{\tilde{\lambda}_k}$, and it peaks when the estimate $\tilde{\lambda}_k$ is closet to $\lambda_i$~\cite{kitaev2002book}. The precision of phase estimation is $2^{-d_t}$, so to ensure that we do not miscount the ground-state degeneracy, we require $2^{-d_t} < a-b$, which implies $d_t = O(\log(n))$ and the largest evolution time is $O(2^{d_t}) = O(\poly(n))$.
We label the best estimate of $\lambda_i$ as $\tilde{\lambda}_i$. Then
\ba
\sum_{k \neq i} \abs{q_k^{(i)}}^2 \leq \delta \ ,
\ea
where $\delta$, the failure probability of phase estimation, is a constant~\cite{kitaev2002book}.

To amplify the success probability, we perform a concatenated phase estimation denoted as $V$~\cite{nagaj_fast_amplification_2009}, which is basically the quantum-circuit version of majority vote. Starting from the state $\left( \ket{0^{d_t}} \right)^{\otimes r} \ket{\psi_i}$, the state after $V$ is given by,
\ba\label{eq:after_concatenated_PE}
\sum_{\vb{k}} q_{\vb{k}}^{(i)} \ket{\tilde{\lambda}_{\vb{k}}} \ket{\psi_i} = \sum_{\vb{j}} q_{\vb{j}}^{(i)} \ket{\tilde{\lambda}_{\vb{j}}} \ket{\psi_i} + \sum_{\vb{l}} q_{\vb{l}}^{(i)} \ket{\tilde{\lambda}_{\vb{l}}} \ket{\psi_i} \ ,
\ea
where $q_{\vb{k}}^{(i)} := q_{k_1}^{(i)} \cdots q_{k_r}^{(i)}$ and $\ket{\tilde{\lambda}_{\vb{k}}} := \ket{\tilde{\lambda}_{k_1}} \cdots \ket{\tilde{\lambda}_{k_r}}$. We have split this summation into two parts. $\ket{\tilde{\lambda}_{\vb{j}}}$ is the state such that the vector $\vb{j}$ has more than a half of its elements equal to~$i$, while the vector $\vb{l}$ has less than a half of its elements equal to~$i$. That is, the subscript $\bf{j}$ corresponds to the success cases and the subscript $\bf{l}$ corresponds to the failure cases. 
By the Chernoff-bound argument, the success probability is amplified to,
\ba
\label{eq:success_probability}
\sum_{\vb{j}} \abs{q_{\vb{j}}^{(i)}}^2 = 1 - 2^{O(-r)} \ .
\ea
By choosing $r = O(\poly(n))$, the occurring probability of the second term in Eq.~\eqref{eq:after_concatenated_PE} is exponentially small, so we can ignore it for simplicity.

Then how can we prepare $\ket{\psi_i}$? It turns out that we do not have to. The trick is the following identity~\cite{yung_metropolis_2012},
\ba
\sum_x \ket{x} \ket{x} = \sum_i \ket{\psi_i} \ket{\psi_i^*} \ ,
\ea
where $\ket{\psi_i^*}$ is the complex conjugate of $\ket{\psi_i}$. So we just need to prepare a maximally entangled state, and then we automatically have all eigenstates of $H$. Applying concatenated phase estimation $V$ to $\ket{0\cdots 0} \sum_x \ket{x} \ket{x}$ yields,
\ba\label{eq:state_after_phase_estimation}
\sum_i \sum_{\vb{j}} q_{\vb{j}}^{(i)} \ket{\tilde{\lambda}_{\vb{j}}} \ket{\psi_i} \ket{\psi_i^*} \ .
\ea

Now, define a function
\ba
f(\tilde{\lambda}) := 
\begin{cases}
0 & \text{if } \tilde{\lambda} \leq b \\
1 & \text{if } \tilde{\lambda} > b
\end{cases} \ .
\ea
Since $f(\tilde{\lambda})$ is a classically efficiently computable Boolean function, we can use a polynomial-sized quantum circuit to evaluate its value, which we defined as $U_f$, i.e. $U_f \ket{0} \ket{\tilde{\lambda}} := \ket{f(\tilde{\lambda})} \ket{\tilde{\lambda}}$. So appending $r$ ancilla qubits to state~\eqref{eq:state_after_phase_estimation} and applying $U_f$ gives,
\ba
\sum_i \sum_{\vb{j}} q_{\vb{j}}^{(i)} \ket{f(\tilde{\lambda}_{\vb{j}})} \ket{\tilde{\lambda}_{\vb{j}}} \ket{\psi_i} \ket{\psi_i^*} \ ,
\ea
where $\ket{f(\tilde{\lambda_{\vb{j}}})}$ is defined in a similar way to $\ket{\tilde{\lambda}_{\vb{j}}}$. Then perform a majority vote $U_{\rm mv}$ to the first $r$ ancillas $\ket{f(\tilde{\lambda}_{\vb{j}})}$ and use another qubit to store the result. Since the vector $\vb{j}$ has more than a half of its elements equal to i, the qubit used to store the result must be in $\ket{f(\tilde{\lambda}_{i})}$; now the state is given by,
\ba
\sum_i \left( \sum_{\vb{j}} q_{\vb{j}}^{(i)} \ket{f(\tilde{\lambda}_{i})} \ket{f(\tilde{\lambda}_{\vb{j}})} \ket{\tilde{\lambda}_{\vb{j}}} \right) \ket{\psi_i} \ket{\psi_i^*} \ .
\ea
After that, we uncompute $\ket{f(\tilde{\lambda}_{\vb{j}})}$ by applying the inverse of $U_f$, which gives,
\ba
\label{eq:before_postselection}
\sum_i \ket{f(\tilde{\lambda}_{i})} \left( \sum_{\vb{j}} q_{\vb{j}}^{(i)} \ket{\tilde{\lambda}_{\vb{j}}} \right) \ket{\psi_i} \ket{\psi_i^*} \ ,
\ea
where we have discarded those qubits reset to $\ket{0\cdots 0}$.

\begin{figure}[t]
\centering
\includegraphics[width=0.45\textwidth]{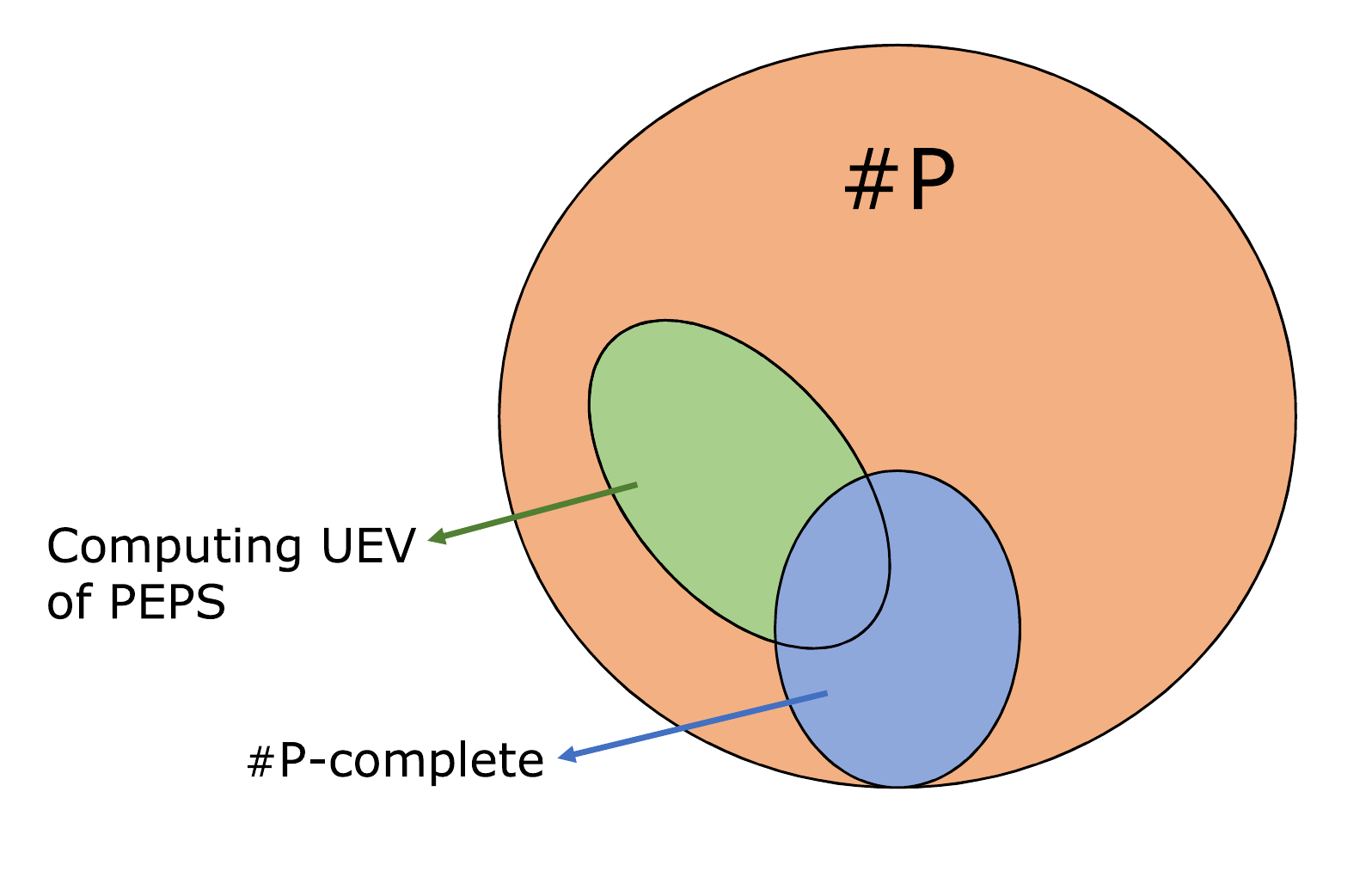}
\caption{Illustration of the fact that computing unnormalized expectation value (UEV) of PEPS is $\sharpP$-complete. Since we have proved that counting ground state degeneracy (CGD) can be reduced to computing UEV of PEPS, CGD is included in the green circle, and thus in $\sharpP$.}
\label{fig:counting_GS_and_sharpP}
\end{figure}

Next, to count the ground-state degeneracy from state~\eqref{eq:before_postselection}, we need to reset the register $\ket{\tilde{\lambda}_{\vb{j}}}$, which can be achieved by applying $V^{\dagger}$. The resulting state is
\ba
\sum_i \ket{f(\tilde{\lambda}_{i})} \left( \sum_{\vb{j}} \abs{q_{\vb{j}}^{(i)}}^2 \right) \ket{0\cdots 0} \ket{\psi_i} \ket{\psi_i^*} + \ket{\text{others}} \ ,
\ea
where in $\ket{\others}$, the second register is not $\ket{0\cdots 0}$ (see Supplemental Materials for details). So post-selecting on the second register being $\ket{0\cdots 0}$ will give the state $\sum_i \left( 1 - 2^{O(-r)} \right) \ket{f(\tilde{\lambda}_{i})} \ket{\psi_i} \ket{\psi_i^*}$, where we have used Eq.~\eqref{eq:success_probability}. The resulting state (unnormalized) is essentially, 
\ba\label{eq:gs_degeneracy_final}
\sum_i \ket{f(\tilde{\lambda}_{i})} \ket{\psi_i} \ket{\psi_i^*} \ .
\ea
The unnormalized expectation value (UEV) of $\dyad{0}\otimes I \otimes I$ with respect to this state is the ground-state degeneracy of $H$.
State~\eqref{eq:gs_degeneracy_final} is generated by a post-selected quantum circuit, so it is also a PEPS~\cite{schuch_PEPS_complexity_2007}. According to Ref.~\cite{schuch_PEPS_complexity_2007}, computing the UEV of a PEPS is $\sharpP$-complete. Therefore, counting ground-state degeneracy is in $\sharpP$ (see Fig.~\ref{fig:counting_GS_and_sharpP} for an illustration).

\textbf{Part II: Counting entanglement spectrum.---}
In this part, we show that counting entanglement spectrum is in $\sharpP$. It is clear that counting entanglement spectrum is equivalent to counting ground state degeneracy if we view $\rho$ as a Hamiltonian. So now the question is how to implement $e^{i\rho t}$ with a post-selected quantum circuit. Our idea is to first encode $\rho$ into a quantum state $\ket{A}$, then implement $\rho$ with $\ket{A}$ as ancillas, and finally use truncated-Taylor-series method~\cite{berry-truncated-taylor} to implement $e^{i \rho t}$.

But there is a technical issue about the gap. Recall that the precision of phase estimation is $O(2^{-d_t})$.
Previously, the spectral gap of the Hamiltonian $H$ is $1/\poly(n)$, which means to distinguish ground states from excited states, the evolution time is $O(2^{d_t}) = O(\poly(n))$. But the `gap' $\Delta_{\lambda}$ of $\rho$ could be exponentially small if the state $\ket{\xi}$ is entangled. So in order for phase estimation to work, the `evolution time' $t$ needs to be $O(2^n)$ and so is the size of the quantum circuit. Nevertheless, as we will see later, as long as the condition $\| \rho t \|_{\infty} = O(\poly(n))$ is satisfied, the size of the phase-estimation circuit is still polynomial. 

\textit{Setp 1: encode $\rho$ into a quantum state $\ket{A}$.---}
$\rho$ is Hermitian, so it can be expanded by Pauli operators, $\rho = \sum_{i=0}^{4^n-1} a_i \sigma_i$, where $\sigma_i$ is tensor product of $I, X, Y, Z$ and $a_0 = 1/2^n$ since $\sigma_0 = I^{\otimes n}$. We can encode the coefficients $a_i$ into a quantum state $\ket{A} := \sum_i a_i \ket{i}$. For details, we refer to Supplemental Materials.

\textit{Step 2: implement $\frac{1}{k!} (i\rho t)^k$.---}
First we will implement $\rho$ with the help of $\ket{A}$. Starting from $\ket{A} \ket{\psi} = \sum_i a_i \ket{i} \ket{\psi}$, we apply $\sigma_i$ to an arbitrary state $\ket{\psi}$ controlled by $\ket{i}$: $\sum_i a_i \ket{i} (\sigma_i \ket{\psi})$. Then apply $\H^{\otimes 2n}$ to $\ket{i}$ followed by post-selection of the first register being $\ket{0\cdots 0}$:
\ba
&\xrightarrow{\H^{\otimes 2n} \text{ on } \ket{i}}& \frac{1}{2^n} \sum_i a_i \ket{0\cdots 0} \left( \sigma_i \ket{\psi} \right) + \ket{\others} \\
&\xrightarrow{ \text{post-selection} }& \frac{1}{2^n} \sum_i a_i \ket{0\cdots 0} \left( \sigma_i \ket{\psi} \right) \notag \\
&=& \frac{1}{2^n} \frac{1}{it} \ket{0\cdots 0} \left( i\rho t \ket{\psi} \right) \, ,
\ea
where, in an abuse of notation, $\ket{\others}$ denotes the superposition of states where the first register is not $\ket{0\ldots 0}$, that is, $\ket{\others}$ is of the form $\sum_{j \neq 0\ldots 0} \ket{j} \ket{\phi_j}$. We might use $\ket{\others}$ to denote similar states in the remaining of the paper. We denote this whole procedure as $V_{\rho t}$ (before post-selection):
\ba
V_{\rho t} \ket{A} \ket{\psi} = \frac{1}{2^n} \frac{1}{it} \ket{0\cdots 0} \left( i\rho t \ket{\psi} \right) + \ket{\others} \ .
\ea

Now, let $\ket{k} := \ket{00\ldots 0 11 \ldots 1}$ be the unary representation of $k$, where the first $K-k$ bits are $0$ and the last $k$ bits are $1$. $K$ is related to the truncated terms in the Taylor expansion of $e^{i\rho t}$. We follow a similar idea of Ref.~\cite{berry-truncated-taylor} to implement $\frac{1}{k!} (i\rho t)^k$, except that now we can use post-selection. First, prepare the state $\ket{k} \ket{0^K} \ket{A}^{\otimes K} \ket{\psi}$. Then applying a unitary $V_{\rm Taylor}$ gives
\ba
\xrightarrow{V_{\rm Taylor}} \ket{k} \ket{0\cdots 0} \left( \frac{(i\rho t)^k}{k!} \ket{\psi} \right) + \ket{k} \ket{\others} \ ,
\ea
where $V_{\rm Taylor}$ comprises controlled-$V_{\rho t}$ and some other controlled rotations. Details about $V_{\rm Taylor}$ can be found in Supplemental Materials. Thus, by post-selecting on the second register being $\ket{0\cdots 0}$, one obtains
\ba
\ket{k} \left( \frac{(i\rho t)^k}{k!} \ket{\psi} \right) \ ,
\ea
where $\ket{0\cdots 0}$ has been discarded .

\textit{Step 3: implement $e^{i\rho t}$.---}
Finally, we are ready to complete the action of $e^{i\rho t}$. We start by creating a uniform superposition of the unary representation $\ket{k}$, which can be achieved by some controlled rotations~\cite{berry-truncated-taylor}. For completeness, we also review the preparation in Supplemental Materials. Then we apply $V_{\rm Taylor}$ and perform post-selection, and the resulting state is given by,
\ba
\sum_{k = 0}^K \ket{k} \left( \frac{(i\rho t)^k}{k!} \ket{\psi} \right) \ .
\ea
After that, we apply $\H^{\otimes K}$ to the first register $\ket{k}$, and post-select on it being $\ket{0^K}$, which gives,
\ba
\sum_{k=0}^K \frac{(i\rho t)^k}{k!} \ket{\psi} \approx e^{i\rho t} \ket{\psi} \ .
\ea
Note that we can implement all above procedures in a unitary way and post-select on all the ancilla qubits being $\ket{0}$ in the end.

In order for the approximation to work, we would require
\ba
\frac{\| (i\rho t)^{K+1} \ket{\psi} \|}{(K+1)!} < \fracp{e \| \rho t\|_{\infty}}{K+1}^{K+1} = O(2^{-\poly(n)}) \ ,
\ea
where the first inequality may be shown using $k! > (k/e)^k$ and $\| A \ket{\psi} \| \leq \| A \|_{\infty}$ for a Hermitian operator $A$. Therefore, $K = O(\| \rho t\|_{\infty}) = O(\poly(n))$ suffices to fulfill the requirement, and the size of the whole circuit is still polynomial.

Now that we have completed the implementation of $e^{i\rho t}$, by setting $t = 2\pi (0, 2, 2^2, \ldots, 2^{d_t})$, we can perform phase estimation. Following the technique of proving the complexity of CGD, we can show that CES is also in $\sharpP$.

\textbf{Part III: $\sharpP$-hardness.---}
Now we are going to prove that both CGD and CES are $\sharpP$-hard. Together with the fact that they are in $\sharpP$, we conclude that these two problems are $\sharpP$-complete. Since $\sharpsat$, the problem of counting satisfying assignments of 2-CNF (Conjuctive Normal Form) formula, is $\sharpP$-complete~\cite{Valiant1979-sharp-2sat}, our idea is to reduce $\sharpsat$ to our problems (with a polynomial-time classical algorithm).

\textit{Reduce $\sharpsat$ to CGD.---}
Given an assignment $x=(x_1, x_2, \ldots, x_n)$ with $x_i \in \{0, 1 \}$, the $k$-CNF formula is ANDs of $\poly(n)$ number of clauses, and each clause contains ORs of $k$ variables which are $x_i$ or $\overline{x_i}$ (NOT $x_i$). Here is an example of 2-CNF: $f(x) = (\overline{x_1} \lor x_2) \land (x_1 \lor \overline{x_3})$.

We can map a 2-CNF formula into a Hamiltonian
\ba
H = \sum_C \dyad{s_i s_j} \otimes I_{\neq i, j} \ .
\ea
Similar mapping was also used in Ref.~\cite{aharonov_quantum_np_2002}.
Here, the subscript $C$ means `clause' and each term of the Hamiltonian corresponds to a clause in the 2-CNF formula. The index $i$ and $j$ are those that appear in the clause $C$ and $s_i$ and $s_j$ are chosen with the following rules:
\ba
\begin{cases}
x_i \lor x_j  \to & s_i s_j = 00\\
\overline{x_i} \lor x_j  \to & s_i s_j = 10 \\
x_i \lor \overline{x_j}  \to & s_i s_j = 01 \\
\overline{x_i} \lor \overline{x_j}  \to & s_i s_j = 11
\end{cases} \ .
\ea
They are the unsatisfying variables to the clause $C$. Those variables that do not appear in the clause are mapped into the identity $I_{\neq i, j}$. In this way, the corresponding Hamiltonian of the 2-CNF formula in our example is: $H = \dyad{1}\otimes \dyad{0} \otimes I + \dyad{0}\otimes I \otimes \dyad{1}$.

The ground state of such Hamiltonian is the satisfying assignment $\ket{x}$ and the ground state energy is $0$, that is $H\ket{x} = 0$ if $f(x) = 1$. Thus, counting the number of satisfying assignments can be reduced to CGD of $H$, implying that CGD is $\sharpP$-hard.

\textit{Reduce CGD to CES of BQPS and PEPS.---}
From now on, we give a subscript $k$ to the $k$-th clause. So $C_k(x) = 1$ means $x$ satisfies $C_k$ and $0$ otherwise. Note that for an assignment $\ket{x}$, the corresponding energy is the number of clauses that it does not satisfy. Then we may write $H = \sum_x N_x \dyad{x}$, where $N_x := |\{ C_k: C_k(x) = 0 \}|$ is the number of unsatisfied clauses of $x$.

$H$ is positive, so we can define a density matrix $\rho$ out of it by $\rho := H/\Tr(H)$. The trace of $H$ is $2^{n-2} \# C$, where $\# C$ is the number of clauses in $f(x)$. It can be seen that the largest eigenvalues of $H$ is $\leq \#C$, so the largest eigenvalue of $\rho$ is no larger than $\lambda^* := 1/2^{n-2}$. The energy gap of $H$ is 1, so the gap in the entanglement spectrum of $\rho$ is $1/2^{n-2}\#C$. Since $\# C = O(\poly(n))$, the gap of $\rho$ matches that of Definition~\ref{def:counting_entanglement_spectrum}. Finally, CGD of $H$ can be reduced to CES above $\lambda^*/\# C$. Therefore, all remains is to prove that $\rho$ is the reduced state of a BQPS, which is also a special case of PEPS.

Consider the following state
\ba
\ket{\xi} = a \sum_s \ket{s}_1 \ket{s}_2 \sum_{k: C_k(s)=0} \ket{k}_3 \ ,
\ea
where $a$ is a normalization factor. Recall that $C_k(s) = 0$ means $C_k$ is not satisfied by $s$. $\ket{k}$ here is the unary representation of $k$. This state is actually a purification of $\rho$: $\Tr_{23}\left[\dyad{\xi} \right] = a^2 H = \rho$ if $a = 1/\sqrt{\Tr(H)}$.
Next, we want to show $\ket{\xi}$ can be generated by a post-selected quantum circuit, so that it is a PEPS~\cite{schuch_PEPS_complexity_2007}. First, prepare the state 
\ba
\frac{1}{\sqrt{2^n \# C}} \sum_s \ket{s} \ket{s} \sum_k \ket{k} \ket{0} \ .
\ea
The first part of this state is a maximally entangled state and the second part is a uniform superposition of $\ket{k}$. Both parts can be prepared efficiently. Then we apply the unitary $U_{\rm eval}$: $U_{\rm eval} \ket{s} \ket{k} \ket{0} = \ket{s} \ket{k} \ket{C_k(s)}$, to evaluate $C_k(s)$ and store it in the last register. Such evaluation can be done classically efficiently, so $U_{\rm eval}$ can also be performed in quantum polynomial time. 
The resulting state is then given by, 
\ba
\frac{1}{2} \ket{\xi}\ket{0} + \frac{\sqrt{3}}{2} \ket{\xi^{\perp}}\ket{1} \ ,
\ea
where $\ket{\xi^{\perp}}$ is the normalized version of $\sum_s \sum_{k: C_k(s)=1} \ket{s} \ket{s} \ket{k}$. In summary, we have a unitary $U$, such that $U \ket{0} \ket{0} = \frac{1}{2} \ket{\xi}\ket{0} + \frac{\sqrt{3}}{2} \ket{\xi^{\perp}}\ket{1}$. Now, we can just post-select on the last qubit being $\ket{0}$ to get $\ket{\xi}$, which implies $\ket{\xi}$ is a PEPS.

But since the amplitude of $\ket{\xi}$ is $1/2$, we can also use oblivious amplitude amplification (Lemma~3.6, Ref.~\cite{berry_exponential_2014}) to amplify the amplitude of $\ket{\xi}$ to $1$, in a unitary way. Concretely, define $U_{\xi} := - U (I\otimes Z) U^{\dagger} (I\otimes Z) U$, then
\ba
U_{\xi} \ket{0}\ket{0} = \ket{\xi} \ket{0} \ .
\ea
Thus, $\ket{\xi}$ is not only a PEPS, but also a BQPS.

In summary, we have proved that CES of a state generated by a polynomial-time quantum circuit (which is a special PEPS) is $\sharpP$-hard.

\begin{figure}[t]
\centering
\includegraphics[width = 0.5\textwidth]{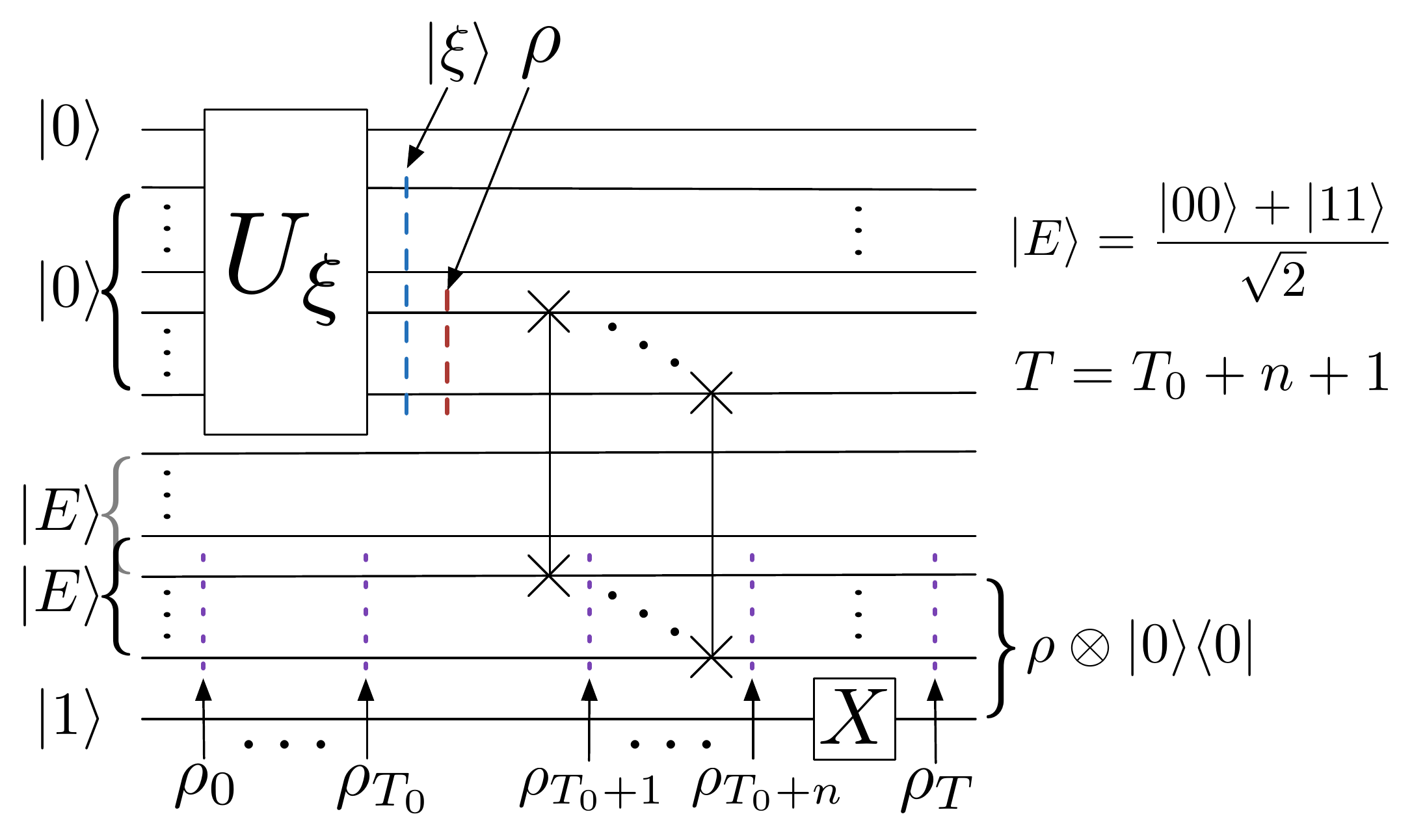}
\caption{History-state construction. There are $n$ EPR pairs $\ket{E}$ in the input. The number of elementary gates in $U_{\xi}$ is $T_0$, which is a polynomial of $n$. The gates between $t = T_0$ and $t = T_0 + n$ are SWAP gates. We are interested in the reduced state of the purple dotted lines $\rho_t$ for $0 \leq t \leq T$.}
\label{fig:history}
\end{figure}

\textit{Reduce CES of PEPS to counting that of G5LS.---}
A gapped 5-local Hamiltonian is of the form $H = \sum_i H_i$ with each $H_i$ acting on at most 5 sites and an inverse polynomial spectral gap. As before, we will first construct a density matrix, and then prove that it is the desired state for our purpose. Consider the density matrix $\tau$, which has the following form,
\ba
\frac{1}{O(\poly(n))} \rho \otimes \dyad{0} + \left( 1 - \frac{1}{O(\poly(n))} \right) \rho' \otimes \dyad{1} \ ,
\ea
where $\rho = H/(2^{n-2} \# C)$ is the reduced state of $\ket{\xi}$ as defined previously. There are following required properties for $\rho'$:
\begin{itemize}
    \item its smallest non-zero eigenvalue is at least $\frac{\lambda^*}{O(\poly(n))}$;
    \item its largest eigenvalue is at most $\lambda^* O(\poly(n))$.
\end{itemize}
Then it can be easily verified that the gap $\Delta_{\lambda}'$ and the largest eigenvalue $\lambda_{\max}'$ of $\tau$ has a difference at most a polynomial factor from $\rho$. Therefore, within the same precision, counting the entanglement spectrum of $\rho$ above $\Delta_{\lambda}$ can be reduced to counting that of $\tau$ above $\Delta'_{\lambda}$.

Now we want to show that $\tau$ is the reduced density matrix of the ground state of a gapped 5-local Hamiltonian. The idea is to use the history-state construction~\cite{kitaev2002book,aharonov_quantum_np_2002}. Consider Fig.~\ref{fig:history}, we make the following claims:
\begin{itemize}
    \item for every $\rho_t$, the largest eigenvalue is at most $O(\poly(n)) \lambda^*$;
    \item there is at least one $\rho_t$ with $t < T$, whose non-zero smallest eigenvalue is at least $\lambda^*/O(\poly(n))$.
\end{itemize}
We prove them in the Supplemental Materials. The history state for Fig.~\ref{fig:history} is
\ba
\ket{\xi'} = \frac{1}{\sqrt{T+1}} \sum_t U_t\cdots U_1 \ket{\alpha} \otimes \ket{t}_C \ ,
\ea
where $\ket{\alpha}$ is the input state to the circuit in Fig.~\ref{fig:history} and $\ket{t}_C$ is the clock state in unary representation. For $t \leq T_0$, $U_t$ is the elementary gate components of $U_{\xi}$, for $T_0 < t \leq T_0 + n$, $U_t$ is a SWAP gate, and for $t = T$, $U_t = X$. The reduced state of the last $n+1$ system qubits of $\ket{\xi'}$ has the following form,
\ba
\frac{1}{T+1} \rho\otimes \dyad{0} + \left(1 - \frac{1}{T+1} \right) \rho' \otimes \dyad{1} \ ,
\ea
if we defined $\rho' := (\sum_{t<T} \rho_t)/T$. Now we verify whether $\rho'$ satisfies those two properties. The eigenvalues of $\rho'$ is greater than eigenvalues of $\rho_0/T$, whose smallest eigenvalue is $1/(2^n T) = \lambda^*/O(\poly(n))$. On the other hand, since for every $\rho_t$, the largest eigenvalue is at most $O(\poly(n)) \lambda^*$, the largest eigenvalue of $\rho'$ is also at most $O(\poly(n)) \lambda^*$.

$\ket{\xi'}$ is the ground state of the following Hamiltonian
\ba
H' = H_{\rm in} + H_{\rm out} + \sum_{t=1}^T H_{\rm prop}(t) + H_{\rm clock} \ ,
\ea
with ground state energy $0$.
The construction of these terms is similar to that of Ref.~\cite{aharonov_quantum_np_2002}, and we leave details to the Supplemental Materials. The relevant facts here are that $H'$ is a 5-local Hamiltonian, and that the four terms of $H'$ has the second smallest eigenvalue at least $\frac{1}{2(T+1)^2}$, which implies the spectral gap of $H'$ is at least $\frac{1}{2(T+1)^2}$, an inverse polynomial of $n$. Thus, we conclude that CES of G5LS is $\sharpP$-complete.

\textbf{Discussion.---}
In this work, we have proved that CES of BQPS, G5LS and PEPS are all $\sharpP$-complete, despite the increasing representational power of these states. A natural question is to ask, for the general tensor-network state, which is a generalization of PEPS, is CES still $\sharpP$-complete or does it belong to a higher complexity class? We leave this question for future research.

Another interesting question is to explore whether such a hardness result holds for the $k$-local-Hamiltonian case with $k < 5$. In the development of the complexity class $\mathsf{QMA}$, it was first proved that 5-local Hamiltonian problem is $\mathsf{QMA}$-complete~\cite{kitaev2002book}, and then shown that the hardness result remains for 2-local Hamiltonian problem~\cite{kempe_complexity_2004}, using the technique of perturbation theory. It might be helpful to modify such a technique to prove that counting entanglement spectrum of ground state of 2-local Hamiltonian is $\sharpP$-complete. But the problem is that the perturbation-theory method is designed for preserving the spectral properties of a Hamiltonian, instead of the entanglement properties of its ground state. We leave as an open problem to extend our result to the 2-local case.

On the other hand, due to the various applications of using entanglement spectrum to characterize many-body systems, it would be interesting to explore the physical implication of our results.

\textbf{Acknowledgement.---}
We acknowledge Xun Gao for helpful discussions.
MHY is supported by the National Natural Science Foundation of China (11875160), the Guangdong Innovative and Entrepreneurial Research Team Program (2016ZT06D348), Natural Science Foundation of Guangdong Province (2017B030308003), and Science, Technology and Innovation Commission of Shenzhen Municipality (ZDSYS20170303165926217, JCYJ20170412152620376, JCYJ20170817105046702).

\bibliography{ref}

\newpage

\clearpage

\newpage
\onecolumngrid

\section{Supplemental Materials}

\beginsupplement

\subsection{Details about the state $\ket{\others}$}

In the main text, when we prove that counting ground state degeneracy is in $\sharpP$, we have the following intermediate state,
\ba
\label{eq:before_V_dagger}
\sum_i \ket{f(\tilde{\lambda}_{i})} \left( \sum_{\vb{j}} q_{\vb{j}}^{(i)} \ket{\tilde{\lambda}_{\vb{j}}} \right) \ket{\psi_i} \ket{\psi_i^*} \ .
\ea
After that, we apply the inverse of concatenated phase estimation $V^\dagger$, which gives,
\ba\label{eq:after_V_dagger}
\sum_i \ket{f(\tilde{\lambda}_{i})} \left( \sum_{\vb{j}} \abs{q_{\vb{j}}^{(i)}}^2 \right) \ket{0\cdots 0} \ket{\psi_i} \ket{\psi_i^*} + \ket{\text{others}} \ ,
\ea
where in $\ket{\others}$, the second register is not $\ket{0\cdots 0}$. In this section, we give a detailed derivation of the above state. 

Suppose we have a unitary $B$, whose action is:
\ba
B\ket{0} = \sum_{j} \beta_j \ket{j} \ ,
\ea
where $j$ is in binary representation. Since $\mel{0}{B^\dagger}{j} = \mel{j}{B}{0}^* = \beta_j^*$, 
\ba
B^\dagger \ket{j} = \beta_j^* \ket{0} + \sum_{j \neq 0} c_j \ket{j} \ .
\ea
That is, the effect of $B^\dagger$ acting on $\ket{j}$ is to give a factor $\beta_j^*$ to $\ket{0\cdots 0}$.

Now, recall that when we apply concatenated phase estimation $V$ to the maximally entangled state, we obtain the following state,
\ba
\sum_i \sum_{\vb{j}} q_{\vb{j}}^{(i)} \ket{\tilde{\lambda}_{\vb{j}}} \ket{\psi_i} \ket{\psi_i^*} \ ,
\ea
which means $q_{\vb{j}}^{(i)}$ plays the role of $\beta_j$. Therefore, when we apply $V^\dagger$ to state~\eqref{eq:before_V_dagger}, it gives,
\ba
\sum_i \ket{f(\tilde{\lambda}_{i})} \left( \sum_{\vb{j}} \abs{q_{\vb{j}}^{(i)}}^2 \right) \ket{0\cdots 0} \ket{\psi_i} \ket{\psi_i^*} + \sum_i \ket{f(\tilde{\lambda}_{i})} \left( \sum_{j \neq 0} c_j \ket{j} \right) \ket{\psi_i} \ket{\psi_i^*} \ .
\ea
Define 
\ba
\ket{\others} := \sum_i \ket{f(\tilde{\lambda}_{i})} \left( \sum_{j \neq 0} c_j \ket{j} \right) \ket{\psi_i} \ket{\psi_i^*} \ ,
\ea
and then we obtain state~\eqref{eq:after_V_dagger}.

\subsection{Encode a density matrix into a quantum state with post-selection}

In this section, we describe how to encode a density matrix $\rho = \sum_{i=0}^{4^n-1} a_i \sigma_i$ into a quantum state $\ket{A}$, where $\sigma_i$ is tensor product of $I, X, Y, Z$ and $a_0 = 1/2^n$. Starting from the state $\ket{0\cdots 0} \ket{\xi}$, we apply $\H^{\otimes n}$ to the first register to prepare $\sum_i \ket{i} \ket{\xi}$. Recall that $\rho$ is the reduced state of $\ket{\xi}$. 
Then apply controlled-$\sigma_i$ to the second register $\ket{\xi}$ controlled by $\ket{i}$, which gives $\sum_i \ket{i} [(I \otimes \sigma_i) \ket{\xi}]$. Finally, post-selecting on the second register being $\ket{\xi}$ gives the state $\ket{A} = \sum_i a_i \ket{i}$, since
\ba
\sum_i \ket{i} \mel{\xi}{I\otimes \sigma_i}{\xi} &=& \sum_i \Tr[(I \otimes \sigma_i) \dyad{\xi} ] \ket{i} \\
&=& \sum_i \Tr[\sigma_i \rho] \ket{i} \\
&=& 2^n \sum_i a_i \ket{i} \ .
\ea

Next, we show how to perform such post-selection. 
Since $\ket{\xi}$ is a PEPS or the ground state of a local Hamiltonian, we can always prepare it using post-selected circuits~\cite{schuch_PEPS_complexity_2007}. We denote the preparation as $V_{\xi}$: 
\ba
V_{\xi} \ket{0} \ket{0} = c \ket{0} \ket{\xi} + \ket{1} \ket{\text{other state}} \ .
\ea
Above is the state before post-selection and $c$ can be exponentially small. We now use our old trick, that is, first applying $V_{\xi}^{\dagger}$ and then post-selecting on the state being $\ket{0}$. The state $(I \otimes \sigma_i) \ket{\xi}$ has two components (up to a normalization factor):
\ba
(I \otimes \sigma_i) \ket{\xi} = a_i \ket{\xi} + \ket{\xi_i^{\perp}} \ ,
\ea
since $\mel{\xi}{I\otimes \sigma_i}{\xi} = \Tr[\sigma_i \rho] = 2^n a_i$.
Then applying $V_{\xi}^\dagger$ to $\ket{0} [(I\otimes \sigma_i)\ket{\xi}]$ gives
\ba
a_i V^{\dagger}_{\xi} \ket{0} \ket{\xi} + V^{\dagger}_{\xi} \ket{0} \ket{\xi_i^{\perp}} \ .
\ea
Since
\ba
\mel{0,0}{V^\dagger_{\xi}}{0, \xi} &=& \mel{0, \xi}{V_{\xi}}{0, 0}^* = c^* \\
\mel{0,0}{V^\dagger_{\xi}}{0, \xi_i^{\perp}} &=& \mel{0, \xi_i^{\perp}}{V_{\xi}}{0, 0}^* = 0 \ ,
\ea
post-selecting on the state being $\ket{0, 0}$ gives $a_i c^* \ket{0,0}$. The whole state (after discarding $\ket{0, 0}$) is $\sum_i a_i \ket{i} = \ket{A}$, which is exactly what we want. To recap, the whole procedure is as follows,
\ba
\ket{0\cdots 0} \ket{0} \ket{\xi} &\xrightarrow{\H^{\otimes n}}& \sum_i \ket{i} \ket{0} \ket{\xi} \\
&\xrightarrow{\text{controlled-}\sigma_i}& \sum_i \ket{i} \ket{0} [(I\otimes \sigma_i) \ket{\xi}] \\
&\xrightarrow{V_{\xi}^{\dagger}}& \sum_i \ket{i} \left( a_i V^{\dagger}_{\xi} \ket{0} \ket{\xi} + V^{\dagger}_{\xi} \ket{0} \ket{\xi_i^{\perp}} \right) \\
&\xrightarrow{\text{post-select on the last two registers being $\ket{0\cdots 0}$}}& c^* \ket{A} \ket{0\cdots 0} + \ket{\others} \ ,
\ea
where $\ket{\others} = \sum_{j \neq 0\ldots 0} \ket{\phi_j} \ket{j}$.


\subsection{Construction of $V_{\rm Taylor}$}

In this section, we show how to implement $\frac{1}{k!} (i\rho t)^k$, specifically, how to construct $V_{\rm Taylor}$. Recall that $\ket{k} = \ket{00\ldots 0 11 \ldots 1}$, where the first $K-k$ bits are $0$ and the last $k$ bits are $1$, and $K$ is related to the truncated terms in the Taylor expansion of $e^{i\rho t}$. First, we prepare $\ket{k} \ket{A}^{\otimes K} \ket{\psi}$. Then for $m = 1, \cdots, K$, apply $V_{\rho t}$ to the $m$-th copy of $\ket{A}$ iff the $m$-th bit of $\ket{k}$ is 1, and we obtain,
\ba
&& \fracp{1}{it}^K (it)^{K-k} k! \fracp{1}{2^n}^k \ket{k} \ket{A}^{K-k} \ket{\bold{0}}^k \left( \frac{(i\rho t)^k}{k!} \ket{\psi} \right) \\
&\xrightarrow[\text{ being $\ket{00\cdots 0}$}]{\text{post-select on $\ket{A}$}}& \fracp{1}{it 2^n}^K (it)^{K-k} k! \ket{k} \ket{00\cdots 0}^K \left( \frac{(i\rho t)^k}{k!} \ket{\psi} \right) \ .
\ea

Now the question is how to eliminate the unwanted factors $(it)^{K-k}$ and $k!$. Define two kinds of rotations:
\ba
V_{\rm rot 1} &:& \ket{0} \to \frac{1}{m} \ket{0} + \sqrt{1-\frac{1}{m^2}} \ket{1} \text{ ($m = 1, \ldots, K$)} \\
V_{\rm rot 2} &:& \ket{0} \to \frac{1}{it} \ket{0} + \sqrt{1+\frac{1}{t^2}} \ket{1} \ .
\ea
Prepare the state $\ket{k} \ket{0^K}$, and then apply $V_{\rm rot 1}$ to the $m$-th bit of $\ket{0^K}$ iff the $m$-th bit of $\ket{k}$ is 1 for $m = 1, \cdots, K$, which gives,
\ba
\frac{1}{k!} \ket{k} \ket{0^K} + \ket{k}\ket{\others} \ ,
\ea
which allows us to eliminate $k!$ after post-selection of the second register being $\ket{0^K}$. Subsequently, we apply $V_{\rm rot 2}$ to the $m$-th bit of $\ket{0^K}$ iff the $m$-th bit of $\ket{k}$ is 0, and the resulting state is given by,
\ba
\frac{1}{k! (it)^{K-k}} \ket{k} \ket{0^K} + \ket{k}\ket{\others} \ .
\ea
So now we can cancel the unwanted factor $k! (it)^{K-k}$. To recap, we can pack all above unitary operations into one and denote it as $V_{\rm Taylor}$. The whole procedure is as follows,
\ba
\ket{k} \ket{0^K} \ket{A}^{\otimes K} \ket{\psi} &\xrightarrow{V_{\rm Taylor}}& \ket{k} \ket{00\cdots 0} \left( \frac{(i\rho t)^k}{k!} \ket{\psi} \right) + \ket{k} \ket{\others} \\
&\xrightarrow[\text{and discard}]{\text{post-select}}& \ket{k} \left( \frac{(i\rho t)^k}{k!} \ket{\psi} \right)
\ea

\subsection{Preparation of uniform superposition of unary representation $\ket{k}$}

To prepare uniform superposition of unary representation $\ket{k}$, we can first apply a rotation to the last qubit, and then apply a controlled rotation to the $i$-th qubit, conditioned on the $(i+1)$-th qubit being $\ket{1}$. Below is a 3-qubit example,
\ba
\ket{000} &\xrightarrow{\text{first rotation}}& \ket{00} (\alpha_1 \ket{0} + \beta_1 \ket{1}) \\
&=& \alpha_1 \ket{00} \ket{0} + \beta_1 \ket{00} \ket{1} \\
&\xrightarrow{\text{second controlled rotation}}& \alpha_1 \ket{000} + \beta_1 \ket{0} (\alpha_2 \ket{0} + \beta_2 \ket{1}) \ket{1} \\
&=& \alpha_1 \ket{000} + \beta_1 \alpha_2 \ket{001} + \beta_1 \beta_2 \ket{011} \\
&\xrightarrow{\text{third controlled rotation}}& \alpha_1 \ket{000} + \beta_1 \alpha_2 \ket{001} + \beta_1 \beta_2 (\alpha_3 \ket{0} + \beta_3 \ket{1}) \ket{11} \\
&=& \alpha_1 \ket{000} + \beta_1 \alpha_2 \ket{001} + \beta_1 \beta_2 \alpha_3 \ket{011} + \beta_1 \beta_2 \beta_3 \ket{111} \ .
\ea
By adjusting the parameters $\alpha_i$ and $\beta_i$, we obtain 
\ba
\frac{1}{\sqrt{K+1}} \sum_{k=0}^K \ket{k} \ .
\ea

\subsection{Details for proving the $\sharpP$-hardness of counting entanglement spectrum of ground state of gapped 5-local Hamiltonians}

We first prove those two claims we made in the main text. For reference, we collect them in the following:
\begin{itemize}
    \item for every $\rho_t$, the largest eigenvalue is at most $O(\poly(n)) \lambda^*$;
    \item there is at least one $\rho_t$ with $t < T$, whose smallest eigenvalue is at least $\lambda^*/O(\poly(n))$.
\end{itemize}
\begin{proof}
Recall that $\rho_t$ is the intermediate reduced density matrix of the qubits indicated by purple dotted lines of Fig.~\ref{fig:history}. The second claim is easier to prove. For $t = 0$, $\rho_0 = I/2^n$, whose smallest eigenvalue is $1/2^n$. The second claim holds since $\lambda^* = 1/2^{n-2}$. 
For the first cliam, note that \textbf{(a)} for $t \leq T_0$, $\rho_t = I/2^n$, so their largest eigenvalue is $\leq \lambda^*$. \textbf{(b)} For $t = T_0 + m$ with $m \leq n$, $\rho_t = \rho_{\xi}^{(m)} \otimes \rho_E^{n-m}$. Here, $\rho_{\xi}^{(m)}$ is the reduced state of the first $m$ qubits of $\rho$ and $\rho_E^{(n-m)} = I/2^{n-m}$ is the reduced state of the last $n-m$ EPR pairs. So $\| \rho_E^{(n-m)} \|_{\infty} = 1/2^{n-m}$. As for $\| \rho_{\xi}^{(m)} \|_{\infty}$, recall that $\rho = H/\left(2^{n-2} \# C \right)$ by construction and $H = \sum_{C_k} \dyad{s_i s_j} \otimes I_{\neq i, j}$. Each term in $H$ is diagonal with diagonal elements at most $1$, and if we trace out $n-m$ qubits, we will get a factor at most $2^{n-m}$. So the largest eigenvalue of $\rho_{\xi}^{(m)}$ is
\ba
\left\| \rho_{\xi}^{(m)} \right\|_{\infty} \leq \sum_{C_k} \frac{2^{n-m}}{2^{n-2} \# C} = \frac{1}{2^{m-2}} \ .
\ea
Then $\| \rho_t \|_{\infty} = \| \rho_{\xi}^{(m)} \|_{\infty} \cdot \| \rho_E^{(n-m)} \|_{\infty} = 1/2^{n-2} \leq O(\poly(n)) \lambda^*$. \textbf{(c)} For $t = T$, $\rho_t = \rho$ and the claim obviously holds.
\end{proof}

Now, we are going to present the concrete form of the four terms in $H'$. First, recall that,
\ba
H' = H_{\rm in} + H_{\rm out} + \sum_{t=1}^T H_{\rm prop}(t) + H_{\rm clock} \ ,
\ea
Then,
\begin{itemize}
    \item The first $2n + \# C + 1$ qubits are the input of $U_{\xi}$, which are all in $0$ state. Let $\ket{E^{\perp}} := \left( \ket{E_1} + \ket{E_2} + \ket{E_3} \right)/\sqrt{3}$, where $\ket{E_i}$ are the other 3 Bell bases. Then
    \ba
    H_{\rm in} = \left( \sum_{i=1}^{2n + \# C + 1} \dyad{1}_i + \sum_{n \text{ EPR pairs}} \dyad{E^{\perp}} + \dyad{0}_{m+2n+1} \right) \otimes \dyad{0}_C \ .
    \ea
    Actually, in the construction of the Hamiltonian, the last register is not in unary representation, and the reason is to make $H'$ a five local Hamiltonian. See Sec.~6 of Ref.~\cite{aharonov_quantum_np_2002} for details.

    \item $H_{\rm out} = \dyad{1}_1 \otimes \dyad{T}_C$

    \item $H_{\rm prop}(t) = \frac{1}{2} \left(I\otimes \dyad{t}_C + I\otimes \dyad{t-1}_C - U_t\otimes \dyad{t}{t-1} - U_t^{\dagger} \otimes \dyad{t-1}{t} \right)$, where $U_t$ is the $t$-th elementary gate component of $U_{\xi}$ for $1 \leq t \leq T_0$, SWAP gate for $T_0 + 1 \leq t \leq T_0 + n$, and Pauli-$X$ for $t = T$.

    \item 
    \ba
    H_{\rm clock} = \sum_{t=1}^T \dyad{01}_{t-1, t} \ ,
    \ea
    where $t$ in the subscript means the $t$-th qubit of the clock state. 
\end{itemize}
It may be verified that under such construction, $\ket{\xi'}$ is the ground state of $H'$ with ground state energy $0$.

\end{document}